\documentclass[10pt]{article}
\usepackage{multirow}
\usepackage{calc}
\usepackage{epsfig,amsfonts,amsthm,latexsym,amsmath}
\usepackage{commath}
\usepackage{algorithm}
\usepackage{algpseudocode}
\usepackage[round]{natbib}

\def\best{\mathit{BestS}}
\def\cbest{\overline{\mathit{BestS}}}

\title{\normalsize\bf%
New algorithms for the Minimum Coloring Cut Problem
}

\author{%
Augusto Bordini$^{1,2}$ \ and \ F\'abio Protti$^{1*}$
}

\begin{document}

\date{}

\maketitle

\vspace{-20pt}
\begin{center}
{\footnotesize
*Corresponding author\\
$^1$Fluminense Federal University\\
Niter\'oi, RJ - Brazil \\
$^2$PETROBRAS\\
Rio de Janeiro, RJ - Brazil\\
E-mails: gutocnet@ic.uff.br / fabio@ic.uff.br
}\end{center}

\bigskip
\noindent
{\small{\bf ABSTRACT.}
The Minimum Coloring Cut Problem is defined as follows: given a connected graph $G$ with colored edges, find an edge cut $E'$ of $G$ (a minimal set of edges whose removal renders the graph disconnected) such that the number of colors used by the edges in $E'$ is minimum. In this work, we present two approaches based on Variable Neighborhood Search to solve this problem. Our algorithms are able to find all the optimum solutions described in the literature.
}


\baselineskip=\normalbaselineskip

\section{Introduction}\label{sec:1}

\begin{figure}
  \centering
  \includegraphics[width=8cm]{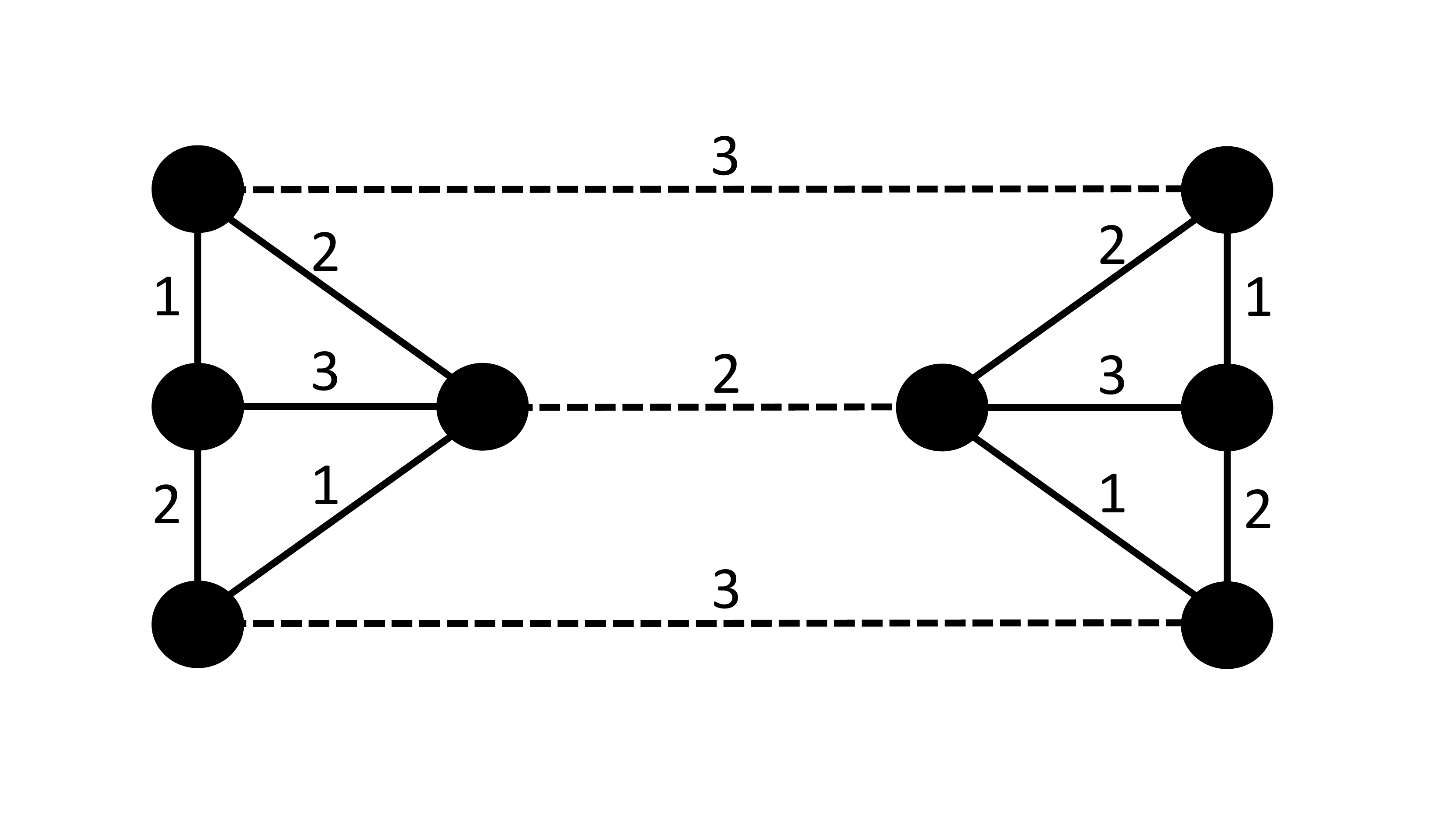}
  \caption{In the above graph, colors are represented by labels in the set $\{1,2,3\}$. The cut consisting of the dashed edges is an optimal solution of the MCCP. The value of the optimal solution is $2$ because the removal of any subset of edges with only one color does not disconnect the graph.}\label{figure1}
\end{figure}

The Minimum Coloring Cut Problem (MCCP) has as input a connected (undirected) graph $G=(V,E)$, with colored (or labeled) edges. Each color is assigned to one or more edges, but each edge $e$ has a unique color $c(e)$. The aim of the MCCP is to find an edge cut $E'$ of $G$ (a minimal set $E'$ of edges such that $G'=(V,E\backslash E')$ is disconnected) with the following property: the set of colors used by the edges in $E'$ has minimum size. Formally:

{\bf Minimum Coloring Cut Problem (MCCP)}\\
{\it Input:} a connected (undirected) graph $G = (V,E,C)$ such that $V$ is the set of nodes of $G$, $E$ is the set of edges of $G$, and $C=\{c(e) \mid e\in E\}$ is the set of colors (or edge labels).\\
{\it Goal:} Find a subset $E'\subseteq E$ such that $G'= (V,E\backslash E')$ is disconnected and the set of colors $C'=\{c(e) \mid e\in E'\}$ is minimized. Figure~\ref{figure1} shows a simple example.

Note that if all the edge colors are distinct then the MCCP amounts to finding a usual minimum cut, a task that can be easily performed in polynomial time using max-flow algorithms. However, the complexity of the MCCP still remains as a theoretical open question. Intuitively, the MCCP is unlikely to be solvable in polynomial-time, because the related problem of finding an $s$-$t$ cut with the minimum number of colors is NP-hard~\citep{Coudert2007}. This fully justifies the design of heuristic algorithms to solve the MCCP.

Colored cut problems are related to the vulnerability of multilayer networks since they provide tight lower bounds on the number of failures that can disconnect totally or partially a network~\citep{Coudert2016}.

The Minimum Color $s$-$t$ Cut Problem (MCstCP for short) is closely related to the MCCP. The input of the MCstCP consists of a connected edge-colored graph $G=(V,E)$ and two nodes $s,t\in V$, and its objective is to find the minimum number of colors whose removal separates $s$ and $t$ in the remaining graph (where `removing a color' means removing all the edges with that color). \citet{Coudert2007} considered the MCstCP for the first time; they prove its NP-hardness and present approximation hardness results. However, five years before, \citet{Jha2002} had already observed that the MCstCP is NP-hard via a simple reduction from the Minimum Hitting Set Problem.

The papers by~\citet{Coudert2007} and~\cite{Coudert2016} approach the MCCP and the MCstCP with the goal of measuring the network's capability of remaining connected when sets of links share risks. For instance, in a WiFi network, an attacker could drop all links on a certain frequency by adding a strong noise signal to it. Other example happens when two links use the same physical environment.

Another potential application of the MCCP is in transportation planning systems, where nodes represent locations served by bus and edge colors represent bus companies. In this case, a solution of the MCCP gives the minimum number of companies that must stop working in order to create pairs of locations not reachable by bus from one another. Such application is more suitably modeled by allowing a multigraph as the input of the MCCP, since two locations can be connected by bus services offered by more than a single company.

\citet{Zhang2014} shows that the MCCP can be solved in polynomial time when the input graph is planar, has bounded treewidth, or has a small value of \emph{fmax} (the maximum number of edges a color is assigned).

In~\citep{Silva2016}, exact methods to solve the MCCP are presented. The authors propose three different integer programming formulations over which branch-and-cut and branch-and-bound approaches are developed. To evaluate their algorithms, they use the instances generated by~\citet{Cerulli2005}.

In some sense, the MCCP is the dual of the Minimum Labelling Spanning Tree Problem (MLSTP), which aims at finding a minimum set $C'$ of colors such that the edges with colors in $C'$ form a connected, spanning subgraph $H$ of $G$. For information on the MLSTP, we refer the reader to~\citep{Krumke1998} and~\citep{Consoli2015}. Note that any spanning tree $T$ of $H$ contains $|C'|$ colors, and thus is a spanning tree of $G$ using a minimum number of colors, i.e., a solution of the MLSTP with input $G$. An analogous argument can be applied to the MCCP: one can first find a disconnecting set $E'$ of edges (not necessarily a cut) that uses a minimum number of colors, and then easily return a minimal disconnecting set $E''\subseteq E'$ as the solution of the MCCP.

Another way of viewing the MCCP is: find a maximum set $C'$ of colors such that $G'=(V,E')$ is disconnected, where $E'=\{e\in E \mid c(e)\in C'\}$, and then pick all the colors in the complementary set $C\backslash C'$. Such strategy is employed by the two new algorithms proposed in this work. The algorithms try to include new colors to the set of current colors, so that adding the edges with those new colors to the current subgraph still keeps it disconnected. When no new color can be included in this way, the colors in $C\backslash C'$ correspond to a solution of the MCCP. Our algorithms are based on the Variable Neighborhood Search (VNS) metaheuristic~\citep{Hansen1997}. As we shall see, the former algorithm uses a greedy, deterministic approach to choose new colors to be included to the current set of colors, while the latter uses a probabilistic approach.

The remainder of this work is structured as follows. In Section~\ref{sec:2} we describe in detail all the functions and procedures used in our algorithms. Section~\ref{sec:3} presents the computational results, where we compare the quality of the solutions obtained by our algorithms with the ones produced by the exact methods described in~\citep{Silva2016}. Section~\ref{sec:4} contains our concluding remarks.

\section{Description of the algorithms}\label{sec:2}

In this section we first describe the general algorithm (Algorithm 1) which is the basic structure for both the greedy, deterministic approach (``VNS-Greedy'') and the probabilistic approach (``VNS-Probabilistic''). Next, we describe in detail each of its subroutines. Some subroutines ( Generate-Initial-Solution, New-Solution, and Local-Search) have a ``greedy version'' and a ``probabilistic version''. Running Algorithm 1 using the greedy versions of such subroutines produces the VNS-Greedy algorithm, while running it using the probabilistic versions produces the VNS-Probabilistic algorithm. The remaining subroutines are common to both approaches.

The description of the general algorithm is as follows:

\newpage

\begin{algorithm}
\caption{General algorithm}\label{alg:general}
\textbf{Input:} Graph $G=(V,E,C)$, where $C=\{c(e)\mid e\in E\}$
\begin{algorithmic}[1]
 \State {Generate-Initial-Solution$(\best)$}
 \State {$\mathit{MaxNeighborhood} \leftarrow \abs{C} - \abs{\best}$}
 \Repeat
   \State {New-Solution$(S)$}
   \While {$\abs{S} > \abs{\best}$}
     \State {$\best \leftarrow S$}
     \State {$\mathit{MaxNeighborhood} \leftarrow \abs{C} - \abs{BestS}$}
     \State {New-Solution$(S)$}
   \EndWhile
   \State {$k \leftarrow 1$}
   \While {$k < \mathit{MaxNeighborhood}$}
     \State {$S' \leftarrow S$}
     \State {Shake$(S',k)$}
     \If {Number-of-Components$(S')=1$}
       \State {Fix$(S')$}
     \EndIf
     \State {Local-Search$(S')$}
     \If {$\abs{S'} > \abs{S}$}
       \State {$S \leftarrow S'$}
       \State {$k \leftarrow 1$}
     \Else { $k \leftarrow k+1$}
     \EndIf
   \EndWhile
   \If {$\abs{S} > \abs{\best}$}
     \State {$\best \leftarrow S$}
     \State {$\mathit{MaxNeighborhood} \leftarrow \abs{C} - \abs{\best}$}
   \EndIf
 \Until {stop condition is true}
 \State {Output the number of colors in the disconnecting set obtained: $\abs{C} - \abs{\best}$}
\end{algorithmic}
\end{algorithm}

Along the execution of the algorithm, a {\em solution} is any subset $C'\subseteq C$ of colors. Let $G'=(V,E')$ be the spanning subgraph of $G$ such that $E'=\{e\in E \mid c(e)\in C'\}$. As an abuse of terminology, we say that solution $C'$ is {\em disconnected} (resp., {\em connected}) if $G'$ is disconnected (resp., connected). Also, we may refer to the number of connected components of $C'$ to mean the number of connected components of $G'$.  The value (number of colors) of solution $C'$ is denoted by $\abs{C'}$. As mentioned in the introduction, we follow the strategy of finding a maximum disconnected solution. To be consistent with this approach, $C'$ is a {\it feasible solution} if and only if $G'$ is disconnected. The complementary set of colors $C\backslash C'$ is denoted by $\overline{C'}$ and called {\em complementary space} of solution $C'$.

Below we discuss the notation used in Algorithm 1:
\begin{itemize}
\item $\best$ is the current best solution. In line 29, the returned value $\abs{C}-\abs{\best}$ is the number of colors in the disconnecting set consisting of all the edges whose colors are in $\cbest$.
\item $\mathit{MaxNeighborhood}$ is a variable that controls the neighborhoods (see line 11) in the core of the VNS strategy (lines 10 to 23).
\item $S$ and $S'$ are auxiliary solutions, explained later.
\item Number-of-Components$(S')$ (line 14) is a standard function that returns the number of connected components of solution $S'$. It is implemented using the well-known disjoint-set (or union-find) data structure with weighted-union heuristic and path compression. Details can be found in~\citep[chapter 21]{Cormen}.
\end{itemize}

An initial solution $\best$ is generated in line 1; next, $\mathit{MaxNeighborhood}$ is set as the number of colors not in $\best$ (line 2). The main loop (lines 3 to 28) is executed until the stop condition is met. The stop condition (maximum running time) is defined empirically according to the instance size (number of nodes $|V|$). After some initial tests, we obtained the values shown in Table 1 below.

\begin{table}[H]
\centering
\label{runtime}
\caption{Stop condition according to instance sizes.}
\begin{tabular}{c|c}
\hline%
number of nodes & max running time (s)\\
\hline
50 & 1\\
100 & 20\\
200 & 30\\
400 & 80\\
500 & 200\\
1000 & 2800\\
\hline
\end{tabular}
\end{table}

In lines 4 to 9, a new candidate solution $S$ is generated in the beginning of a new iteration. First, $S$ is generated using subroutine New-Solution. (line 4). If $S$ is better than $\best$ then $\best$ and $\mathit{MaxNeighborhood}$ are updated and another candidate solution $S$ is generated by New-Solution. The {\bf while} loop (lines 5 to 9) ends when the number of colors of the candidate solution is not greater than the number of colors of the current best solution.

Lines 10 to 23 contain the core of the basic VNS strategy~\citep{Hansen1997}. For each candidate solution $S$, $S'$ is set to $S$ (line 12), and then the shaking and local search procedures are executed over $S'$ for $k$ iterations, where $k$ controls the neighborhoods and ranges in $1\,.\,.\,\mathit{MaxNeighborhood}$. If shaking and local search are able to improve $S'$ so that $\abs{S'}>\abs{S}$ then $S$ is updated and $k$ is restarted to $1$, i.e., a new cycle of $k$ iterations begins.

When $k$ is equal to $\mathit{MaxNeighborhood}$, the current best solution $\best$ is compared with $S$ and updated if necessary (lines 24 to 27). The execution stops if the maximum running time is reached (line 28); otherwise, it returns to the candidate solution generation step.

When the stop condition is true, the value $\abs{C}-\abs{\best}$ is returned. The subset of edges $E'=\{e\in E\mid c(e)\in\cbest\}$ is a disconnecting set using $\abs{C}-\abs{\best}$ colors. If needed, a cut can be obtained by finding any minimal disconnecting set $E''\subseteq E'$.

In the next subsections we describe in detail the subroutines used in Algorithm 1. When applicable, the greedy and probabilistic versions of a subroutine are presented.

\subsection{Generate-Initial-Solution}

This subroutine has a greedy version (Algorithm 2) and a probabilistic version (Algorithm 3). In the greedy version, the initial solution is constructed iteratively color by color. At each step, a color $c$ not appearing in the current solution is greedily chosen so that the number of connected components of $\best\cup\{c\}$ is maximized. The subroutine stops when every color in the complementary set $\cbest$ turns the current solution connected when added to it.

\begin{algorithm}
\caption{Generate-Initial-Solution$(\best)$ -- greedy version}
\label{alg:initial-solution-greedy}
\begin{algorithmic}[1]
 \State {$\best \leftarrow \emptyset$}
 \State {$\mathit{endloop} \leftarrow \mathit{false}$}
 \Repeat
   \State {let $c\in \cbest$ be a color maximizing Number-of-Components$(\best\cup\{c\})$}
   \If {Number-of-Components$(\best\cup\{c\}) > 1$}
     \State {$\best \leftarrow \best\cup\{c\}$}
   \Else { $\mathit{endloop} \leftarrow \mathit{true}$}
   \EndIf
 \Until {$\mathit{endloop} = \mathit{true}$}
\end{algorithmic}
\end{algorithm}

Adding a color that maximizes the number of connected components (line 4 in Algorithm 2) usually guides the subroutine to locally optimal solutions. This strategy is precisely the deterministic approach used by~\citet{Krumke1998} and other authors for the  MLSTP.

To avoid local optima, we use an adapted Boltzmann function that allows a probabilistic color choice at each iteration. Such adapted Boltzmann function is inspired by the Simulated Annealing Cooling Schedule described in~\citep{Aarts2005}, and is used not only in subroutine Generate-Initial-Solution, but also in subroutines New-Solution and Local-Search.

We remark that the probabilistic versions of subroutines Generate-Initial-Solution, New-Solution and Local-Search differ from the greedy ones precisely in the choice strategy of colors to be included in the current best solution.

The probability $P(c)$ of a color $c$ to be included in the current best solution $\best$ is directly proportional to the number of connected components of $\best\cup\{c\}$. Let $\gamma\in \cbest$ be the color that maximizes Number-of-Components$(\best\cup\{\gamma\})$. The probabilities $P(c)$ are normalized by the Boltzmann function values $\exp(\Delta(c)/T)$, where:

$\bullet$ $\Delta(c) =$ Number-of-Components$(\best\cup\{c\})$ $-$ Number-of-Components$(\best\cup\{\gamma\})$

$\bullet$ $T$ is a parameter referred to as {\em temperature} that controls the function's dynamic; in our experiments we use $T=1$.

\begin{algorithm}
\caption{Generate-Initial-Solution$(\best)$ -- probabilistic version}
\label{alg:initial-solution-probabilistic}
\begin{algorithmic}[1]
 \State {$\best \leftarrow \emptyset$}
 \State {$\mathit{endloop} \leftarrow \mathit{false}$}
 \Repeat
   \State {let $\gamma\in \cbest$ be a color maximizing Number-of-Components$(\best\cup\{\gamma\})$}
   \For {\textbf{each} $c\in \cbest$}
     \State {determine the probability $P(c)$ normalized by Boltzmann function $\exp(\Delta(c))$}
   \EndFor
   \If{there is a color $c\in \cbest$ such that $\best\cup\{c\}$ is feasible}
        \State{following the probabilities $P(\,)$, randomly select a color $c\in \cbest$\\
        \hspace*{1.1cm} such that $\best\cup\{c\}$ is feasible}
     \State {$\best\leftarrow\best\cup\{c\}$}
   \Else { $\mathit{endloop} \leftarrow \mathit{true}$}
   \EndIf
 \Until {$\mathit{endloop} = \mathit{true}$}
\end{algorithmic}
\end{algorithm}

\subsection{New-Solution}

New-solution is a subroutine used to generate a candidate solution $S$ at the beginning of a new iteration in the {\bf repeat} loop (lines 3 to 28) of Algorithm 1. It is implemented as a local search~\citep{Hansen1997} on the colors in $\cbest$ as an attempt to raise the diversity factor, since the complementary space of $\best$ is a completely different search zone with respect to the current best solution.

Algorithms 4 and 5 are, respectively, the greedy and probabilistic versions of subroutine New-Solution. Our tests revealed that both algorithms produce an immediate peak of diversification as the local search evolves.

In order to extract a feasible solution from $\cbest$ an iterative process of inclusion of new colors is performed as follows.

Solution $S$ is initialized as an empty set of edges (line 1 in both algorithms). Note that the number of connected components of $S$ at this moment is $|V|$ (corresponding to a spanning subgraph containing only isolated vertices).

The first {\bf while} loop (lines 2 to 8 in Algorithm 4, and 2 to 13 in Algorithm 5) generates a partial solution $S$ color by color, and stops in two cases:

(a) the set $\cbest\backslash S$ of unused colors is empty;

(b) every remaining color in $\cbest\backslash S$ would generate an infeasible (connected) solution if added to current solution $S$.

The second {\bf while} loop (lines 9 to 15 in Algorithm 4, and 14 to 25 in Algorithm 5) works in the same way, but try to add to current solution $S$ colors from $\best$ instead. It stops when no color in $\best\backslash S$ is able to produce a feasible solution when added to $S$.

\begin{algorithm}
\caption{New-Solution$(S)$ -- greedy version}
\label{alg:new-solution-greedy}
\begin{algorithmic}[1]
 \State {$S \leftarrow \emptyset$}
 \While {Number-of-Components$(S)>1$ \textbf{and} $\cbest\backslash S\neq\emptyset$}
   \State {let $c\in\cbest\backslash S$ be a color maximizing Number-of-Components$(S\cup\{c\})$}
   \If {Number-of-Components$(S\cup\{c\})>1$}
     \State {$S\leftarrow S\cup\{c\}$}
   \Else { break}
   \EndIf
 \EndWhile
 \While {Number-of-Components$(S)>1$}
   \State {let $c\in\best\backslash S$ be a color maximizing Number-of-Components$(S\cup\{c\})$}
   \If {Number-of-Components$(S\cup\{c\})>1$}
     \State {$S\leftarrow S\cup\{c\}$}
   \Else { break}
   \EndIf
 \EndWhile
\end{algorithmic}
\end{algorithm}

\begin{algorithm}
\caption{New-Solution$(S)$ -- probabilistic version}
\label{alg:new-solution-probabilistic}
\begin{algorithmic}[1]
 \State {$S \leftarrow \emptyset$}
 \While {Number-of-Components$(S)>1$ \textbf{and} $\cbest\backslash S\neq\emptyset$}
   \State {let $\gamma\in\cbest\backslash S$ be a color maximizing Number-of-Components$(S\cup\{\gamma\})$}
   \For {\textbf{each} $c\in \cbest\backslash S$}
     \State {determine the probability $P(c)$ normalized by Boltzmann function $\exp(\Delta(c))$}
   \EndFor
   \If{there is a color $c\in \cbest\backslash S$ such that $S\cup\{c\}$ is feasible}
        \State{following the probabilities $P(\,)$, randomly select a color $c\in \cbest\backslash S$\\
        \hspace*{1.1cm} such that $S\cup\{c\}$ is feasible}
     \State {$S\leftarrow S\cup\{c\}$}
   \Else { break}
   \EndIf
 \EndWhile
 \While {Number-of-Components$(S)>1$}
   \State {let $\gamma\in\best\backslash S$ be a color maximizing Number-of-Components$(S\cup\{\gamma\})$}
   \For {\textbf{each} $c\in \best\backslash S$}
     \State {determine the probability $P(c)$ normalized by Boltzmann function $\exp(\Delta(c))$}
   \EndFor
   \If{there is a color $c\in \best\backslash S$ such that $S\cup\{c\}$ is feasible}
        \State{following the probabilities $P(\,)$, randomly select a color $c\in \best\backslash S$\\
        \hspace*{1.1cm} such that $S\cup\{c\}$ is feasible}
     \State {$S\leftarrow S\cup\{c\}$}
   \Else { break}
   \EndIf
 \EndWhile
\end{algorithmic}
\end{algorithm}

\subsection{Shake}

This subroutine is common to both VNS-Greedy and VNS-Probabilistic. It consists of finding a new solution by adding/removing $k$ colors randomly from current solution $S'$, in order to diversify the range of solutions and try to escape from a local optimum. The total number of operations (additions plus removals) depends on $k$ (parameter passed from the main algorithm), which is the {\em size} of the neighborhood. The value of $k$ ranges from $1$ to the maximum neighborhood size (variable $\mathit{MaxNeighborhood}$). 

In line 2, $\delta$ is a random value in $[0,1]$. In line 3, it is necessary to check whether $\abs{S'}>0$ before removing a color from $S'$. At the end of Algorithm 6, the symmetric difference between solutions $S$ and $S'$ contains exactly $k$ colors, i.e., $\abs{(S\backslash S')\cup(S'\backslash S)}=k$.

We remark that, after the shaking, the new solution $S'$ may be infeasible (connected). The purpose of subroutine Fix (explained in the next subsection) is to deal with such event.

\begin{algorithm}
\caption{Shake$(S',k)$}
\label{alg:shaking}

{\bf Input:} solution $S'$ and size of neighborhood $k$

\begin{algorithmic}[1]
 \For {$i=1,\ldots,k$}
   \State $\delta\leftarrow\mathit{random}(0,1)$  
   \If {$\delta < 0.5$ \textbf{and} $\abs{S'}>0$}
     \State {randomly remove a color from $S'\cap S$}
   \Else { randomly add a color $c\in \overline{S'}\cap\overline{S}$ to $S'$}
   \EndIf
 \EndFor
\end{algorithmic}
\end{algorithm}

\subsection{Fix}

This subroutine is also common to VNS-Greedy and VNS-Probabilistic. If after the shaking procedure $S'$ is infeasible (line 14 in Algorithm 1), subroutine Fix is invoked. It consists of iteratively removing colors at random from $S'$ until it turns into a feasible solution.

\begin{algorithm}
\caption{Fix$(S')$}
\label{alg:fix}
\begin{algorithmic}[1]
 \While {Number-of-Components$(S') = 1$}
   \State {randomly remove a color from $S'$}
 \EndWhile
\end{algorithmic}
\end{algorithm}

\subsection{Local-Search}

The subroutine Local-Search has a greedy version (Algorithm 8) and a probabilistic version (Algorithm 9).

In the greedy version, after solution $S'$ is submitted to subroutines Shake and Fix, new colors are greedily added to $S'$ until no longer possible.

The probabilistic version is similar, but the choice of new colors follows the strategy already described in the probabilistic versions of subroutines Generate-Initial-Solution (Algorithm 3) and New-Solution (Algorithm 5).

\begin{algorithm}
\caption{Local-Search$(S')$ -- greedy version}
\label{alg:local-search-greedy}
\begin{algorithmic}[1]
 \While {Number-of-Components$(S')> 1$}
   \State {let $c\in\overline{S'}$ be the color that maximizes Number-of-Components$(S'\cup\{c\})$}
   \If {Number-of-Components$(S'\cup\{c\})> 1$}
     \State {$S'\leftarrow S'\cup\{c\}$}
   \Else { break}
   \EndIf
 \EndWhile
\end{algorithmic}
\end{algorithm}

\begin{algorithm}
\caption{Local-Search$(S')$ -- probabilistic version}
\label{alg:local-search-probabilistic}
\begin{algorithmic}[1]
 \While {Number-of-Components$(S')>1$}
   \State {let $\gamma\in\overline{S'}$ be a color maximizing Number-of-Components$(S'\cup\{\gamma\})$}
   \For {\textbf{each} $c\in\overline{S'}$}
     \State {determine the probability $P(c)$ normalized by Boltzmann function $\exp(\Delta(c))$}
   \EndFor
   \If{there is a color $c\in\overline{S'}$ such that $S'\cup\{c\}$ is feasible}
        \State{following the probabilities $P(\,)$, randomly select a color $c\in\overline{S'}$\\
        \hspace*{1.1cm} such that $S'\cup\{c\}$ is feasible}
     \State {$S'\leftarrow S'\cup\{c\}$}
   \Else { break}
   \EndIf
 \EndWhile
\end{algorithmic}
\end{algorithm}

\section{Computational Results}\label{sec:3}

The experiments were performed on an Intel Core I7 4GHz with 32Gb RAM, running Linux Ubuntu x64 14.04 operating system. Algorithms were implemented in C++ and compiled using optimization flag -O3.

Our experiments were performed using the $720$ problem instances created by~\citet{Cerulli2005}, divided in $72$ datasets containing $10$ randomly generated instances each. All the $10$ instances in a single dataset have the same number of nodes $|V|$, number of colors $|C|$, and edge density $d$; that is, each dataset is characterized by a prescribed triple $(|V|,|C|,d)$. The {\em expected} number of edges $|E|$ of an instance is $d\,|V|\,(|V|-1)/2$; thus, in a same dataset, instances may have slightly different values of $|E|$. The value of $|V|$ ranges in the set $\{50,100,200,400,500,1000\}$, while the value of $d$ in $\{0.2,\,0.5,\,0.8\}$ (corresponding, respectively, to a low, medium, or high density). The value of $|C|$ varies according to the instance size. For example, if $|V|=50$ then $|C|\in\{12,25,50,62\}$. Tables 2 to 7 show all the combinations $(|V|,|C|,d)$ used in our tests. Each row in a table corresponds to the $10$ instances of a single dataset. For each dataset, solution quality is evaluated as the average solution value (number of colors in the solution) calculated over the $10$ problem instances.

Maximum allowed CPU times were chosen as stop conditions for the algorithms, determined according to instance sizes (see Table 1 in Section 2.1).

In Tables 2 to 7, our results are compared with the results obtained by the three exact methods proposed in~\citep{Silva2016}. In all the tables, the first and second columns show, respectively, the number of colors and the density; in the third column, each entry shows the average solution value obtained by the exact methods over the $10$ instances of the corresponding row (a symbol `-' means that the methods were unable to find the optima); in the fourth column, each entry shows the average computational time of the exact method that best deals with the $10$ instances of the corresponding row (a symbol `-' means that the runs were aborted after reaching a time limit); columns 5 and 6 (resp., 7 and 8) have the same meaning as columns 3 and 4, but refer to our VNS greedy (resp., VNS probabilistic) approach.

For instances with the same number of nodes, the tests show, as expected, that low density instances converge faster than medium/high density instances, because the latter have larger search spaces.

The exact methods proposed in~\cite{Silva2016} are able to find optimum solutions only for $|V|\leq 200$. In this scenario (see Tables 2 to 4), both the VNS greedy and VNS probabilistic approaches reach all the optimum solutions, in lower computational times.

For $|V|\in\{400,500,1000\}$ (see Tables 5 to 7), the VNS greedy and VNS probabilistic approaches found exactly the same average solution value for all datasets. The VNS probabilistic approach is faster for $50$-node instances (see Table 2). For other values of $|V|$ (see Tables 3 to 7), no algorithm clearly outperforms the other in terms of computational times.

\section{Conclusions}\label{sec:4}

In this paper we described new VNS-based algorithms for the MCCP. Previously to this work, no other results for the MCCP besides the ones obtained by~\citet{Silva2016} were known for instances up to $200$ nodes (to the best of the authors' knowledge). Our algorithms reach all the known optimal solutions in lower computational times. For instances with unknown optima, our algorithms provide the same solutions, in reasonable computational times.

Computational experiments were performed using two different approaches, greedy and probabilistic, in order to evaluate how the algorithms are influenced by the color choice strategy. Computational results showed that the two approaches exhibit the same behavior in terms of solution quality, and no significant difference in terms of computational times.

\begin{table}[ht]
\centering
\label{tab:50 nodes}
\caption{Computational results for instances with {\bf 50} nodes}
\begin{footnotesize}
\resizebox{!}{.09\paperheight}{%
\begin{tabular}{ccccccccccc}
\hline%
\multicolumn{2}{c}{\bf Parameters} & {} & \multicolumn{2}{c}{\bf\citet{Silva2016}}& {} & \multicolumn{2}{c}{\bf VNS-Greedy} & {} & \multicolumn{2}{c}{\bf VNS-Probabilistic}\\
    \cline{1-2}
    \cline{4-5}
    \cline{7-8}
    \cline{10-11}
colors & density & {} & value & time (s) & {} & value & time (s) & {} & value & time (s)\\
    \cline{1-11}
    {\multirow{3}{*}{12}} & 0.8 & {} & 9.8 & 0.001 & {} & 9.8 & 0.001 & {} & 9.8 & 0.0004\\
    \multicolumn{1}{c}{}&0.5 & {} & 7.4 & 0.001 & {} & 7.4 & 0.007 & {} & 7.4 & 0.0004\\
    \multicolumn{1}{c}{}&0.2 & {} & 2.5 & 0.006 & {} & 2.5 & 0.003 & {} & 2.5 & 0.0003\\
    \cline{1-11}
    {\multirow{3}{*}{25}} & 0.8 & {} & 15.5 & 0.05 & {} & 15.5 & 0.004 & {} & 15.5 & 0.001\\
    \multicolumn{1}{c}{}&0.5 & {} & 9.9 & 0.04 & {} & 9.9 & 0.009 & {} & 9.9 & 0.001\\
    \multicolumn{1}{c}{}&0.2 & {} & 2.7 & 0.009 & {} & 2.7 & 0.008 & {} & 2.7 & 0.0008\\
    \cline{1-11}
    {\multirow{3}{*}{50}} & 0.8 & {} & 21.3 & 1.48 & {} & 21.3 & 0.01 & {} & 21.3 & 0.006\\
    \multicolumn{1}{c}{}&0.5 & {} & 11.6 & 0.82 & {} & 11.6 & 0.01 & {} & 11.6 & 0.007\\
    \multicolumn{1}{c}{}&0.2 & {} & 2.8 & 0.04 & {} & 2.8 & 0.01 & {} & 2.8 & 0.003\\
    \cline{1-11}
    {\multirow{3}{*}{62}} & 0.8 & {} & 22.7 & 1.8 & {} & 22.7 & 0.02 & {} & 22.7 & 0.007\\
    \multicolumn{1}{c}{}&0.5 & {} & 12.1 & 1.1 & {} & 12.1 & 0.01 & {} & 12.1 & 0.006\\
    \multicolumn{1}{c}{}&0.2 & {} & 2.8 & 0.05 & {} & 2.8 & 0.01 & {} & 2.8 & 0.004\\
    \cline{1-11}
\end{tabular}}
\end{footnotesize}
\end{table}

\begin{table}[ht]
\centering
\label{tab:100 nodes}
\caption{Computational results for instances with {\bf 100} nodes}
\begin{footnotesize}
\resizebox{!}{.09\paperheight}{%
\begin{tabular}{ccccccccccc}
\hline%
\multicolumn{2}{c}{\bf Parameters} & {} & \multicolumn{2}{c}{\bf\citet{Silva2016}} & {} & \multicolumn{2}{c}{\bf VNS-Greedy} & {} & \multicolumn{2}{c}{\bf VNS-Probabilistic}\\
    \cline{1-2}
    \cline{4-5}
    \cline{7-8}
    \cline{10-11}
colors & density & {} & value & time (s) & {} & value & time (s) & {} & value & time (s)\\
    \cline{1-11}
    {\multirow{3}{*}{25}} & 0.8 & {} & 21 & 0.1 & {} & 21 & 0.004 & {} & 21 & 0.003\\
    \multicolumn{1}{c}{}&0.5 & {} & 16.5 & 0.09 & {} & 16.5 & 0.01 & {} & 16.5 & 0.006\\
    \multicolumn{1}{c}{}&0.2 & {} & 6.2 & 0.06 & {} & 6.2 & 0.02 & {} & 6.2 & 0.002\\
    \cline{1-11}
    {\multirow{3}{*}{50}} & 0.8 & {} & 33.1 & 4.8 & {} & 33.1 & 0.02 & {} & 33.1 & 0.03\\
    \multicolumn{1}{c}{}&0.5 & {} & 22.2 & 5.7 & {} & 22.2 & 0.05 & {} & 22.2 & 0.02\\
    \multicolumn{1}{c}{}&0.2 & {} & 6.8 & 0.4 & {} & 6.8 & 0.01 & {} & 6.8 & 0.01\\
    \cline{1-11}
    {\multirow{3}{*}{100}} & 0.8 & {} & 45.2 & 22.4 & {} & 45.2 & 0.1 & {} & 45.2 & 0.09\\
    \multicolumn{1}{c}{}&0.5 & {} & 26.5 & 9.3 & {} & 26.5 & 0.1 & {} & 26.5 & 0.09\\
    \multicolumn{1}{c}{}&0.2 & {} & 7.2 & 2.1 & {} & 7.2 & 0.06 & {} & 7.2 & 0.06\\
    \cline{1-11}
    {\multirow{3}{*}{125}} & 0.8 & {} & $\langle\,45.2\,\rangle$ & $\langle\,22.4\,\rangle$ & {} & 48.6 & 0.1 & {} & 48.6 & 0.1\\
    \multicolumn{1}{c}{}&0.5 & {} & 27.1 & 36.1 & {} & 27.1 & 0.1 & {} & 27.1 & 0.1\\
    \multicolumn{1}{c}{}&0.2 & {} & 7.2 & 3.1 & {} & 7.2 & 0.07 & {} & 7.2 & 0.08\\
    \cline{1-11}
    \multicolumn{11}{l}{{\tiny Obs.: Values between brackets are probably typing errors - they repeat the information given three rows above.}}
\end{tabular}}
\end{footnotesize}
\end{table}

\begin{table}[H]
\centering
\label{tab:200 nodes}
\caption{Computational results for instances with {\bf 200} nodes}
\begin{footnotesize}
\resizebox{!}{.09\paperheight}{%
\begin{tabular}{ccccccccccc}
\hline%
\multicolumn{2}{c}{\bf Parameters} & {} & \multicolumn{2}{c}{\bf\citet{Silva2016}}& {} & \multicolumn{2}{c}{\bf VNS-Greedy} & {} & \multicolumn{2}{c}{\bf VNS-Probabilistic}\\
    \cline{1-2}
    \cline{4-5}
    \cline{7-8}
    \cline{10-11}
colors & density & {} & value & time (s) & {} & value & time (s) & {} & value & time (s)\\
    \cline{1-11}
    {\multirow{3}{*}{50}} & 0.8 & {} & 43.3 & 2.5 & {} & 43.3 & 0.05 & {} & 43.3 & 0.05\\
    \multicolumn{1}{c}{}&0.5 & {} & 32.7 & 9.4 & {} & 32.7 & 0.1 & {} & 32.7 & 0.09\\
    \multicolumn{1}{c}{}&0.2 & {} & 13.2 & 5.7 & {} & 13.2 & 0.08 & {} & 13.2 & 0.07\\
    \cline{1-11}
    {\multirow{3}{*}{100}} & 0.8 & {} & 68.8 & 238.9 & {} & 68.8 & 0.3 & {} & 68.8 & 0.3\\
    \multicolumn{1}{c}{}&0.5 & {} & 45.4 & 699.5 & {} & 45.4 & 0.3 & {} & 45.4 & 0.3\\
    \multicolumn{1}{c}{}&0.2 & {} & 15 & 188.3 & {} & 15 & 0.2 & {} & 15 & 0.2\\
    \cline{1-11}
    {\multirow{3}{*}{200}} & 0.8 & {} & 93.8 & 2051.9 & {} & 93.8 & 1.2 & {} & 93.8 & 1.2\\
    \multicolumn{1}{c}{}&0.5 & {} & 54.1 & 2066.0 & {} & 54.1 & 1.0 & {} & 54.1 & 1.1\\
    \multicolumn{1}{c}{}&0.2 & {} & 15.9 & 614.4 & {} & 15.9 & 1.0 & {} & 15.9 & 1.2\\
    \cline{1-11}
    {\multirow{3}{*}{250}} & 0.8 & {} & $\langle\,93.8\,\rangle$ & $\langle\,2051.9\,\rangle$ & {} & 99.4 & 1.7 & {} & 99.4 & 1.9\\
    \multicolumn{1}{c}{}&0.5 & {} & 56.5 & 2990.2 & {} & 56.5 & 1.7 & {} & 56.5 & 1.7\\
    \multicolumn{1}{c}{}&0.2 & {} & 16.1 & 691.6 & {} & 16.1 & 1.3 & {} & 16.1 & 1.1\\
    \cline{1-11}
    \multicolumn{11}{l}{{\tiny Obs.: Values between brackets are probably typing errors - they repeat the information given three rows above.}}
\end{tabular}}
\end{footnotesize}
\end{table}

\begin{table}[ht]
\centering
\label{tab:400 nodes}
\caption{Computational results for instances with {\bf 400} nodes}
\begin{footnotesize}
\resizebox{!}{.09\paperheight}{%
\begin{tabular}{ccccccccccc}
\hline%
\multicolumn{2}{c}{\bf Parameters} & {} & \multicolumn{2}{c}{\bf\citet{Silva2016}}& {} & \multicolumn{2}{c}{\bf VNS-Greedy} & {} & \multicolumn{2}{c}{\bf VNS-Probabilistic}\\
    \cline{1-2}
    \cline{4-5}
    \cline{7-8}
    \cline{10-11}
colors & density & {} & value & time (s) & {} & value & time (s) & {} & value & time (s)\\
    \cline{1-11}
    {\multirow{3}{*}{100}} & 0.8 & {} & - & - & {} & 88.7 & 0.5 & {} & 88.7 & 0.5\\
    \multicolumn{1}{c}{}&0.5 & {} & - & - & {} & 71.9 & 0.9 & {} & 71.9 & 0.9\\
    \multicolumn{1}{c}{}&0.2 & {} & - & - & {} & 30.7 & 1.1 & {} & 30.7 & 1.1\\
    \cline{1-11}
    {\multirow{3}{*}{200}} & 0.8 & {} & - & - & {} & 144.3 & 3.4 & {} & 144.3 & 5.5\\
    \multicolumn{1}{c}{}&0.5 & {} & - & - & {} & 99.5 & 4.3 & {} & 99.5 & 4.2\\
    \multicolumn{1}{c}{}&0.2 & {} & - & - & {} & 35 & 4.7 & {} & 35 & 4.7\\
    \cline{1-11}
    {\multirow{3}{*}{400}} & 0.8 & {} & - & - & {} & 195.7 & 15.7 & {} & 195.7 & 17.6\\
    \multicolumn{1}{c}{}&0.5 & {} & - & - & {} & 120.3 & 22.6 & {} & 120.3 & 20.3\\
    \multicolumn{1}{c}{}&0.2 & {} & - & - & {} & 37.4 & 11.5 & {} & 37.4 & 14.5\\
    \cline{1-11}
    {\multirow{3}{*}{500}} & 0.8 & {} & - & - & {} & 210.2 & 26.7 & {} & 210.2 & 27.5\\
    \multicolumn{1}{c}{}&0.5 & {} & - & - & {} & 124.9 & 26.6 & {} & 124.9 & 33.0\\
    \multicolumn{1}{c}{}&0.2 & {} & - & - & {} & 38.2 & 19.2 & {} & 38.2 & 20.4\\
    \cline{1-11}
\end{tabular}}
\end{footnotesize}
\end{table}

\begin{table}[ht]
\centering
\label{tab:500 nodes}
\caption{Computational results for instances with {\bf 500} nodes}
\begin{footnotesize}
\resizebox{!}{.09\paperheight}{%
\begin{tabular}{ccccccccccc}
\hline%
\multicolumn{2}{c}{\bf Parameters} & {} & \multicolumn{2}{c}{\bf\citet{Silva2016}}& {} & \multicolumn{2}{c}{\bf VNS-Greedy} & {} & \multicolumn{2}{c}{\bf VNS-Probabilistic}\\
    \cline{1-2}
    \cline{4-5}
    \cline{7-8}
    \cline{10-11}
colors & density & {} & value & time (s) & {} & value & time (s) & {} & value & time (s)\\
    \cline{1-11}
    {\multirow{3}{*}{125}} & 0.8 & {} & - & - & {} & 111.4 & 1.3 & {} & 111.4 & 1.6\\
    \multicolumn{1}{c}{}&0.5 & {} & - & - & {} & 89.2 & 1.4 & {} & 89.2 & 2.5\\
    \multicolumn{1}{c}{}&0.2 & {} & - & - & {} & 37.1 & 2.6 & {} & 37.1 & 2.5\\
    \cline{1-11}
    {\multirow{3}{*}{250}} & 0.8 & {} & - & - & {} & 178.3 & 9.5 & {} & 178.3 & 9.9\\
    \multicolumn{1}{c}{}&0.5 & {} & - & - & {} & 123.8 & 10.3 & {} & 123.8 & 11.7\\
    \multicolumn{1}{c}{}&0.2 & {} & - & - & {} & 41.4 & 12.1 & {} & 41.4 & 8.6\\
    \cline{1-11}
    {\multirow{3}{*}{500}} & 0.8 & {} & - & - & {} & 240.4 & 44.8 & {} & 240.4 & 36.5\\
    \multicolumn{1}{c}{}&0.5 & {} & - & - & {} & 146.8 & 63.5 & {} & 146.8 & 36.4\\
    \multicolumn{1}{c}{}&0.2 & {} & - & - & {} & 45 & 28.5 & {} & 45 & 27.7\\
    \cline{1-11}
    {\multirow{3}{*}{625}} & 0.8 & {} & - & - & {} & 256.9 & 51.7 & {} & 256.9 & 54.8\\
    \multicolumn{1}{c}{}&0.5 & {} & - & - & {} & 155.2 & 55.4 & {} & 155.2 & 72.6\\
    \multicolumn{1}{c}{}&0.2 & {} & - & - & {} & 45.3 & 43.2 & {} & 45.3 & 51.8\\
    \cline{1-11}
\end{tabular}}
\end{footnotesize}
\end{table}

\begin{table}[H]
\centering
\label{tab:1000 nodes}
\caption{Computational results for instances with {\bf 1000} nodes}
\begin{footnotesize}
\resizebox{!}{.09\paperheight}{%
\begin{tabular}{ccccccccccc}
\hline%
\multicolumn{2}{c}{\bf Parameters} & {} & \multicolumn{2}{c}{\bf\citet{Silva2016}}& {} & \multicolumn{2}{c}{\bf VNS-Greedy} & {} & \multicolumn{2}{c}{\bf VNS-Probabilistic}\\
    \cline{1-2}
    \cline{4-5}
    \cline{7-8}
    \cline{10-11}
colors & density & {} & value & time (s) & {} & value & time (s) & {} & value & time (s)\\
    \cline{1-11}
    {\multirow{3}{*}{250}} & 0.8 & {} & - & - & {} & 228.8 & 20.6 & {} & 228.8 & 18.3\\
    \multicolumn{1}{c}{}&0.5 & {} & - & - & {} & 197.2 & 49.5 & {} & 197.2 & 36.1\\
    \multicolumn{1}{c}{}&0.2 & {} & - & - & {} & 113.8 & 71.8 & {} & 113.8 & 67.2\\
    \cline{1-11}
    {\multirow{3}{*}{500}} & 0.8 & {} & - & - & {} & 375.4 & 176.1 & {} & 375.4 & 134.7\\
    \multicolumn{1}{c}{}&0.5 & {} & - & - & {} & 284.3 & 156.2 & {} & 284.3 & 277.5\\
    \multicolumn{1}{c}{}&0.2 & {} & - & - & {} & 133.4 & 223.6 & {} & 133.4 & 188.9\\
    \cline{1-11}
    {\multirow{3}{*}{1000}} & 0.8 & {} & - & - & {} & 514.7 & 559.7 & {} & 514.7 & 645.7\\
    \multicolumn{1}{c}{}&0.5 & {} & - & - & {} & 353.6 & 736.1 & {} & 353.6 & 820.0\\
    \multicolumn{1}{c}{}&0.2 & {} & - & - & {} & 145.8 & 525.4 & {} & 145.8 & 580.8\\
    \cline{1-11}
    {\multirow{3}{*}{1250}} & 0.8 & {} & - & - & {} & 552.6 & 1109.8 & {} & 552.6 & 1189.2\\
    \multicolumn{1}{c}{}&0.5 & {} & - & - & {} & 369.7 & 1230.6 & {} & 369.7 & 1061.8\\
    \multicolumn{1}{c}{}&0.2 & {} & - & - & {} & 147 & 789.1 & {} & 147 & 885.2\\
    \cline{1-11}
\end{tabular}}
\end{footnotesize}
\end{table}

\bibliographystyle{abbrvnat}
\bibliography{mcc}

\medskip
Received xxxxx 2017 / accepted xxx 2017.
\end{document}